\documentclass[sigconf]{acmart}

\usepackage{booktabs} 

\usepackage[inline]{enumitem}
\setlist[description]{leftmargin=0.0em}

\usepackage[linesnumbered,ruled]{algorithm2e}
\usepackage{algorithmic}

\usepackage{color}
\newcommand {\out}[1]{}

\setcopyright{none}

\renewcommand\footnotetextcopyrightpermission[1]{} 
\begin{document}

\title{A Short Note on Proximity-based Scoring of \\ Documents with Multiple Fields}

\author{Tomohiro Manabe}
\affiliation{%
       \institution{Yahoo Japan Corporation}
       \city{Chiyoda-ku, Tokyo}
       \country{Japan}
       \postcode{102-8282}
}
\email{tomanabe@yahoo-corp.jp}

\author{Sumio Fujita}
\affiliation{%
       \institution{Yahoo Japan Corporation}
       \city{Chiyoda-ku, Tokyo}
       \country{Japan}
       \postcode{102-8282}
}
\email{sufujita@yahoo-corp.jp}

\begin{abstract}
The BM25 ranking function is one of the most \out{famous}well known query relevance\out{-biased} document scoring functions and many variations of it are proposed\out{BM25}. The BM25F function is one of its adaptations designed for modeling documents with multiple fields. 
The Expanded Span method extends a BM25-like function by taking into considerations of the proximity between term occurrences. 
In this note, we combine these two variations into one scoring method in view of proximity-based scoring of documents with multiple fields.
\end{abstract}


\maketitle

\section{Introduction} \label{s:intro}

In this note, we propose a scoring function for keyword-based document retrieval,
where document scoring is the most crucial part of the ranked retrieval against a document collection.
We adopt a document representation as an array of term occurrence positions in order to facilitate the extension to proximity based scoring.
Given an adequate document scoring function, retrieval systems sort documents in descending order of their estimated probabilities that their contents are relevant to the user search intents.
State-of-the-art document scoring functions consist of a linear combination of a term weighting function of each query term, computed against the document representation.

The BM25 function \cite{bm25} is one of the most well known keyword-based document scoring functions. 
For more sophisticated modeling of documents, many variations of BM25 have been proposed. 
Specifically, the BM25F ranking function \cite{bm25f} is for documents with multiple fields of different weights, whereas the Expanded Span method \cite{span} takes into account the proximity of term occurrences in documents.

Although both field weighting and term proximity are important and independent evidences for a document being relevant, 
there exist only a few methods which use both evidences as the best of our knowledge.
In this note, we propose a term proximity sensitive scoring function for documents with multiple fields by plugging the Expanded Span method into a variant of BM25F.

\section{Methods}

In this section, we explain the BM25 ranking function, BM25F and Expanded Span variations, succeeded by our method to combine these variations.

\subsection{BM25 ranking function}
The BM25 ranking function \cite{bm25} is a simple approximation to 2-Poisson model term weighting where 
within document term frequencies exhibit either an \emph{eliteness} distribution of term representing the main topic of the document or a \emph{non-elite} distribution of term occurrence used in a general meaning.
Term weighting function represents how much influence the number of term occurrences in a document affects the estimation of document relevance.

Formally, the BM25 score of a document $D$ against a query $Q$ is:
$$\sum_{t \in Q}\frac{(k_1 + 1)\mathrm{tf}(t, D)}{K(D) + \mathrm{tf}(t, D)} \log\frac{N - \mathrm{df}(t) + 0.5}{\mathrm{df}(t) + 0.5} \ \mathrm{,}$$
$$K(D) = k_1 \cdot \{(1 - b) + b \cdot \mathrm{len}(D) / \mathrm{avgLen}\}\ ,$$
where $t$ denotes a term, tf$(t, D)$ the number of occurrences of $t$ in $D$, $N$ the number of all documents in the search target collection, df$(t)$ the number of documents in which $t$ occurs, len$(D)$ the number of all term occurrences in $D$, and avgLen the arithmetic mean of len$(D)$ for all documents. 
Its parameters are $k_1$ ($> 0$) and $b$ (in $[0, 1]$). 
Giving a lower value to $k_1$ results in a flat term weight function, which assigns a higher score to a document with the larger coverage of query terms whereas a infinite large value to $k_{1}$ results in a linear function to the term frequency.
The $b$ parameter adjusts the strength of the term frequency normalization by document length. 
Note that we just ignored duplicated term occurrences in $Q$ and the number of relevant documents 
that is unknown in ad hoc retrieval contexts (i.e., \textit{qtf}, $R$, and $r$ in the original paper \cite{bm25}).

\subsection{BM25F ranking function}
\label{BM25F}
The BM25F ranking function \cite{bm25f} treats structured documents with a set of typed fields.\out{as sets of categorized fields.}
Each field has an array of term occurrences just as a single unstructured document.
Fields are typed and named according to their functionality in the whole document so that we can arrange an adequate processing against each field according to the index, search and other service strategies. 
For example, item name, specification, description, published date and price are possible field names of e-commerce search.
Each field is typed by, such as text string, integer number, date, and so on, but we focus only on text string field in this paper.

The BM25F function is only different from BM25 for its consideration of weighted fields. 
Denoting a field by $f$, the function is:
$$\sum_{t \in Q}\frac{w(t, D)}{k_1 + w(t, D)} \log\frac{N - \mathrm{df}(t) + 0.5}{\mathrm{df}(t) + 0.5} \ \mathrm{, }$$
$$w(t, D) = \sum_{f \in D} \frac{\mathrm{tf}(t, f, D) \cdot boost_{f}}{(1 - b_f) + b_f \cdot \mathrm{len}(f, D) / \mathrm{avgLen}(f)}\ ,$$
where tf$(t, f, D)$ is the number of occurrences of $t$ in field $f$ of document $D$, len$(f, D)$ the number of all term occurrences in the field, and avgLen$(f)$ the arithmetic mean of len$(f, D)$ for all documents.
Of course, we can specify $boost_f$ values which are relative weights of each field.
We can also specify $b_f$ values, which corresponds to BM25 $b$, for each field. 
Note that the original function \cite{bm25f} does not take different $b$ values for different fields, instead we use an extension of BM25F to enable field dependent normalization. 
For example, another paper describing a publicly available search software implementation \cite{bm25f_solr} also referred to such a variation.

\subsection{Expanded Span Method}
The Expanded Span method \cite{span} treats documents as an array of term occurrences like BM25. 
For scoring a document against a given query, the method first extracts expanded spans, i.e., chain of the ordered query term matches in the document, calculate relevance contributions of the expanded spans about each term, 
and then score the document against the query by integrating the relevance contributions upon local and global statistics with a BM25-like function.

\out{Extraction of spans is not a topic of this note.}
We assume that expanded spans are already extracted from documents with the method in the original paper \cite{span}. Thus obtained expanded spans should satisfy the following constraints:
\begin{itemize}
\out{\item Spans are sub-arrays of documents.}
\item A span is a chain of ordered query term occurrences in the document.
\item Spans never overlap to each other.
\item Each query term occurs in a span once at most.
\out{\item Distance between adjoining term occurrences in a span is the value of parameter $M$ (a positive integer) at most.}
\item Maximum distance between adjoining occurrences in a span is limited to parameter $M$ (a positive integer).
\out{\item Both ends of a span are keyword occurrences.}
\end{itemize}

The relevance contribution of spans in a document $D$ against a term $t$ is:
$$\mathrm{rc}(t, D) = \sum_{s \in D} \mathrm{in}(t, s) \cdot \mathrm{len}(s)^{z} / \mathrm{width}(s)^{x} ,$$
where $s$ denotes a span, in$(t, s)$ is 1 if $t$ occurs in $s$ otherwise 0, and len$(s)$ denotes the number of term positions in $s$.
width$(s)$ is the difference between the first and the last term positions in $s$ or $1 / M$ if width$(s) = 1$.
The parameters are $z$, $x$, and $M$. The values of $z$ and $x$ adjust the impact of the term frequency in a span and the span width respectively. 
$z$ corresponds to $x + y$ in the original paper \cite{span}.
The score is integrated into a document scoring function by replacing tf$(t, D)$ in BM25 by rc$(t, D)$.

\subsection{Combination of BM25F and Expanded Span}

Our key ideas are independent calculation of relevance contribution of each field and relevance contribution integration based on a BM25F-like function.
Our method treats a document as a set of named fields like BM25F as described in Section \ref{BM25F}. 
Formally, relevance contribution of spans in the $f$ field of document $D$ against a term $t$ is:

$$\mathrm{rc}(t, f, D) = \sum_{s \in D[f]} \mathrm{in}(t, s) \cdot \mathrm{len}(s)^{z_f} / \mathrm{width}(s)^{x_f} ,$$
where $D[f]$ denotes the $f$ field of $D$, $s$ an expanded span. 
The $z_f$ and $x_f$ parameters for each field correspond to $z$ and $x$ parameters in the Expanded Span method.
The score is integrated into a document scoring function by replacing tf$(t, f, D)$ in BM25F by rc$(t, f, D)$.

\section{Concluding Remarks for Parameter Calibration}
In this note, we proposed a proximity-based scoring function for documents with multiple fields.
Our document scoring function has $4K+1$ parameters to be calibrated empirically where $K$ is the number of field, as described as follows: 
\begin{itemize}
\item[{\bf $k_{1}$}] As this is the main parameter of the original BM25 function, quite a few papers reported the values used in empirical studies, which are applicable to our proposed method as well.
\item[{\bf $boost_{f}$}] This takes $K-1$ parameters to be tuned just same as BM25F. For shorter and the most important field such as ``title'', as high as 2 to 3 may be given whereas 1 for the ``body'' field. 
Anyway, this parameter seems to be simply collection dependent and to be hard to find a general calibration strategy. Thus, it would be better to use, for example, learn to rank techniques to learn parameters to the target collection by decomposing a document relevance score into field dependent features.
\item[{\bf $b_{f}$}] Unlike the original BM25F function, the proposed function requires $K-1$ more parameters to be tuned. Similar to the case of $k_{1}$, we may refer to past literature. It is also worth noting that the very short field such as ``title'' or constant length field such as ``abstract'' do not seem to require any normalization as they are out of the scope of the original hypothesis behind BM25, where it is enough to set zero to $b_{f}$.
\item[{\bf $x_{f},z_{f}$}] It is easy to suppose that giving larger values to them easily results in a deterioration of search results by overweighting phrasal terms against single terms. In the example of the original paper \cite{span}, the authors set $0.25$ to $x$ and $0.3$ to $y=z-x$ against TREC-9,10 and 11 web track collections, otherwise ``{\it thousand year}'' might be overweighted against ``{\it sea}''. 
\item[{\bf $M$}] The original paper \cite{span} reported that a window size as large as $45$ would be better against the above mentioned collections. They concluded that a better term proximity weighting is able to benefit from a larger window size.
\end{itemize}
An example of automatically calibrating basic BM25 parameters is found in \cite{fujita09ipm}, but still it is challenging to learn a large number of parameters.

\out{
One of its major limitations is that it does not consider proximity between term occurrences in different fields. Solving the problem is one of the future works.}

\bibliographystyle{ACM-Reference-Format}
\bibliography{arxiv17manabe}  


\end{document}